\documentclass[a4paper,11pt]{article}
\usepackage{amsmath,amsfonts,amssymb,amsthm,amstext,amscd}
\usepackage{latexsym}
\usepackage[hidelinks=true]{hyperref}
\usepackage{graphicx}
\usepackage{caption}
\usepackage{comment}
\usepackage{graphicx}
\usepackage{cite}
\usepackage{color}
\usepackage{setspace}
\usepackage{float}
\usepackage[mathscr]{eucal}
\usepackage{xcolor}
\usepackage{titletoc}
\dottedcontents{section}[1.2em]{}{1em}{1pc}

\newcommand{\mrd}{\mathrm{d}}

\definecolor{darkred}{rgb}{0.9,0.05,0.05}
\definecolor{darkblue}{rgb}{0.05,0.05,0.6}
\definecolor{darkgreen}{rgb}{0.05,0.6,0.05}
\definecolor{brightgreen}{rgb}{0.1,0.9,0.1}

\renewcommand*{\eqref}[1]{%
  \begingroup
    \hypersetup{
      linkcolor=darkblue,
      linkbordercolor=darkblue,
    }%
    \textcolor{darkblue}{(\ref{#1})}%
  \endgroup 
}

\hypersetup{linkcolor=red,citecolor=brightgreen,urlcolor=black,colorlinks=true} 
\newcommand*{\secref}[1]{\textcolor{red}{\ref{#1}}}

\numberwithin{equation}{section}

\begin{document}

\setlength{\skip\footins}{1cm}

\begin{titlepage}

\begin{flushright}\vspace{-3.5cm}
{\small
IPM/P-2016/004 \\
February 17, 2016}\end{flushright}
\vspace{0.5cm}

\begin{center}

\parbox{1.05\linewidth}{\fontsize{17pt}{20pt}{\hspace*{-0.28cm}\bf{Conserved Charges and First Law of Thermodynamics}\hspace*{0.65cm}}}\\
\vspace*{0.2cm}
\LARGE\textbf{for Kerr-de Sitter Black Holes}
 \vspace{9mm}

\centerline{\large{Kamal Hajian\footnote{e-mail: kamalhajian@ipm.ir}}}

\vspace{3mm}
\normalsize
 \textit{School of Physics, Institute for Research in Fundamental
Sciences (IPM), \\P.O. Box 19395-5531, Tehran, Iran}
\vspace{5mm}




\begin{abstract}
\noindent
Recently, a general method for calculating conserved charges for (black hole) solutions to generally covariant gravitational theories, in any dimensions and with arbitrary asymptotic behaviors has been introduced. Equipped with this method, which can be dubbed as ``solution phase space method," we calculate mass and angular momentum for the Kerr-dS black holes. Furthermore, for any choice of horizons, associated entropy and the first law of thermodynamics are derived. Interestingly, according to insensitivity of the analysis to the chosen cosmological constant, the analysis unifies the thermodynamics of rotating stationary black holes in 4 (and other) dimensions with either AdS, flat or dS asymptotics. We extend the analysis to include electric charge, i.e. to the Kerr-Newman-dS black holes.
\end{abstract}
\vspace*{0.0cm}
\end{center}
\footnotesize{$\ast$ The final publication is available at Springer via \href{http://dx.doi.org/10.1007/s10714-016-2108-4}{http://dx.doi.org/10.1007/s10714-016-2108-4.}}

\vspace*{0.4cm}

\renewcommand{\baselinestretch}{1.05}  
\setstretch{1.5}
\hypersetup{linkcolor=black,citecolor=darkgreen,urlcolor=darkgreen,colorlinks=true} 
\textcolor{black}{\tableofcontents}
\hypersetup{linkcolor=darkred,citecolor=darkgreen,urlcolor=darkgreen,colorlinks=true} 
\end{titlepage}

\setlength{\skip\footins}{1cm}

\section{Introduction and summary of the results}\label{sec Intro}
In the early 70s,  black holes (BHs) were distinguished to show thermodynamic behaviors. By a semi-classical analysis, their temperature (which is called Hawking temperature) was shown to be related to their geometry by $T_{_\mathrm{H}}=\frac{\kappa}{2\pi}$ where $\kappa$ is the surface gravity on their event horizon \cite{Hawking:1976rt}. Seeking analogues of the first and second laws of thermodynamics to hold for the BHs, and with the help of the Hawking temperature, their entropy was also read from their geometry. Specifically for the Einstein-Hilbert (EH) gravity, the entropy (which is called Bekenstein-Hawking entropy) was found to be $S=\frac{A_{_\mathrm{H}}}{4G}$, in which $A_{_\mathrm{H}}$ is the area of the event horizon  \cite{Bekenstein:1973ft,Bardeen:1973gd}. One could also find the same result for the entropy of a black hole, by studying directly the entropy of the Hawking radiation emitted in the whole process of the evaporation \cite{Zurek:1982zz,Aghapour:2016nik}. Nonetheless, a robust classical, semi-classical or quantum description for the microstates corresponding to the origin of this entropy is still an open question, although some interesting proposals have been suggested. One of the famous works in this direction was utilizing the string theory to describe microstates of some supersymmetric BHs \cite{Strominger:1996sh}. Another appreciated attempts for the realization of BH microstates have been based on the loop quantum gravity \cite{Rovelli:1996dv,Ashtekar:1997yu,Meissner:2004ju}.  A seminal progress in describing BH entropy was made in 1993-94 by Iyer and Wald \cite{Wald:1993nt,Iyer:1994ys}, who defined the entropy as a Noether conserved charge associated with the normalized horizon Killing vector, calculated by an integral over the bifurcation surface of the BH event horizon. This definition confirmed the Bekenstein-Hawking entropy for the EH theory. In addition, it made the definition of the entropy independent of the first law.  As a result, the first law of thermodynamics (in the absence of Abelian gauge fields) could be proved as an identity relating variation of different conserved charges \cite{Iyer:1994ys}.

Kerr-de Sitter BHs (Kerr-dS) are a family of BHs in $4$-dimensional spacetime as solutions to the EH gravity with positive cosmological constant \cite{Carter:1970ea,Carter:1970ea2}(see Refs. \cite{Spradlin:2001pw,Akcay:2010vt} as nice reviews). After the realization of the thermodynamic behaviors in BH physics, Kerr-dS BHs have been also studied in this context \cite{Gibbons:1977mu,Abbott:1981ff}. By the discovery of accelerating expansion of the universe \cite{Perlmutter:1998np}, which can be modelled as a de Sitter spacetime, the study of Kerr-dS BHs has found more motivations. Besides, constructing a consistent dS/CFT correspondence has necessitated the understanding of the thermodynamics of asymptotic de Sitter BHs \cite{Strominger:2001pn,Klemm:2001ea,Balasubramanian:2001nb,Ghezelbash:2001vs,Dehghani:2001kn,Dehghani:2002np,
Dehghani:2002nt}. Nonetheless, the literature on this subject is not yet well-established due to the appearance of negative or ambiguous masses, absence of well-defined asymptotics, ambiguity in the choice of Killing vector to which mass is associated, dealing with first laws containing pressure terms, or terms relating entropies of different horizons, and problems in regularization of conserved charges and choosing appropriate and unique reference points for them. In order to have a glance at the literature of Kerr-dS BH's conserved charges and thermodynamics, see \emph{e.g.} Refs.  \cite{Gibbons:1977mu,Abbott:1981ff,Spradlin:2001pw,Klemm:2001ea,Balasubramanian:2001nb,
Ghezelbash:2001vs,Dehghani:2001kn,Dehghani:2002np, Cai:2001tv, Deser:2002jk, Deser:2005jf, Deser:2007vs, Teitelboim:2002cv, Gomberoff:2003ea,
Dehghani:2002nt,Choudhury:2004ph,Ghezelbash:2004af,
Akcay:2010vt,Sekiwa:2006qj,Chrusciel:2013epa,Chrusciel:2015sna,Dolan:2013ft,McInerney:2015xwa,Kubiznak:2015bya}. The goal of this paper is to circumvent the difficulties alluded above, and provide a coherent and firm thermodynamic description for these geometries. 

Recently, a method for calculating conserved charges associated with ``exact symmetries" of (black hole) solutions in generally covariant gravitational theories, for generic asymptotics and in any dimension has been presented \cite{HS:2015xlp}. We can dub it as ``solution phase space method", or SPSM for short.  It casts calculation of conserved charges into a simple unified formulation. This method has enabled us to circumvent difficulties mentioned above based on some of its peculiar properties (which will be discussed thoroughly):
\begin{itemize}
\item[--] Conserved charges associated with ``exact symmetries" (specifically mass, angular momenta and electric charges associated with stationarity, axial isometries and global gauge transformations) can be calculated by an integration over (almost) \emph{arbitrary} co-dimension-2 surfaces. As a result, mass, angular momenta, and electric charges can be considered as charges attributed to the whole geometry, \emph{i.e.} independent of any specific horizon, asymptotics, etc.
\item[--] In SPSM, calculated conserved charges are automatically regular and unambiguous.
\item[--] Entropy is dealt as a conserved charge associated with a specific exact symmetry, on the same footing as other charges. Hence, it can also be calculated over arbitrary codimension-2 surfaces, even in the presence of nonvanishing energy-momentum tensors. 
\item[--] Entropy is a property attributed to a horizon through the definition of its exact symmetry Killing vector field/generator: Its generator would be a linear combination of generators of stationarity, axial isometry and global gauge transformation, in which the linear coefficients are determined by the choice of the horizon. 
\item[--] For any chosen horizon, the coefficients relating the entropy generator to the generators of other conserved charges are exactly the same coefficients appearing in the first law(s), which relate(s) variations of the entropies to the variations of other charges. So, by linearity of the conserved charges in their generators, there would be a ``first law" for each one of the chosen horizons. 
\item[--] The presence of a rigid gauge transformation in the generator of the entropy remedies the deficiency of Iyer-Wald proof of the BH's first law, when there are some gauge fields. 
\end{itemize}

SPSM enables us to calculate mass and angular momentum for Kerr-dS BHs independent of any horizon or asymptotics. Then, for any chosen horizon, entropy and first law can be calculated and proved.  Interestingly, the analysis shows that thermodynamics of these BHs can be explained in a unified way with the Kerr and Kerr-AdS BHs.   The results are as follows. Consider stationary BHs in $4$-dimensional spacetime as solutions to the theory described by the Lagrangian density $\mathcal{L}=\frac{1}{16\pi G} (R-2\Lambda)$, which are asymptotically AdS, flat, or dS (depending on the sign of $\Lambda$). Their asymptotically-non-rotating metric can be written as
\begin{align}\label{general metric}
\hspace{-0.23cm}\mathrm{d}s^2= -&\Delta_\theta(\frac{1-\frac{\Lambda r^2}{3}}{\Xi}\!-\!\Delta_\theta f)\mathrm{d}t^2+\frac{\rho ^2}{\Delta_r}\mathrm{d}r^2+\frac{\rho ^2}{\Delta_\theta} \mathrm{d}\theta ^2 -2\Delta_\theta fa\sin ^2 \theta\,\mathrm{d}t \mathrm{d}\varphi\nonumber \\
+&\left( \frac{r^2+a^2}{\Xi}+fa^2\sin ^2\theta \right)\sin ^2\theta\,\mathrm{d}\varphi ^2\,,
\end{align}
where
\begin{align}
\rho^2 &\equiv r^2+a^2 \cos^2 \theta\,,\qquad \Delta_r \equiv (r^2+a^2)(1-\frac{\Lambda r^2}{3})-2Gmr\,,\nonumber\\
\Delta_\theta&\equiv 1+\frac{\Lambda a^2}{3}\cos ^2\theta\,,\qquad \Xi\equiv 1+\frac{\Lambda a^2}{3}\,,\qquad
f\equiv\frac{2Gmr}{\rho ^2\Xi^2}\,.\nonumber
\end{align}
Mass and angular momentum as conserved charges associated with the exact symmetry generators $\partial_t$ and $\partial_\varphi$ are calculated for these geometries to be unambiguously
\begin{equation}
M=\frac{m}{\Xi^2}\,,\qquad J=\frac{ma}{\Xi^2}\,,
\end{equation}
independent of any chosen horizon or asymptotics. Assuming legitimate parameters $0\leq m$ and $a$, and for any chosen $\Lambda$, these conserved charges are always positive.  On the other hand, entropy, surface gravity and angular velocity associated with each one of the horizons present in the geometry can be expressed respectively as
\begin{align}
 S_{_\mathrm H}=\frac{\pi(r_{_\mathrm{H}}^2+a^2)}{G\Xi}\,, \qquad \kappa_{_\mathrm H}=\frac{r_{_\mathrm H}(1-\frac{\Lambda a^2}{3}-\Lambda{r_{_\mathrm H}^2}-\frac{a^2}{r_{_\mathrm H}^2})}{2(r_{_\mathrm H}^2+a^2)}\,,\qquad \Omega_{_\mathrm H}=\frac{a(1-\frac{\Lambda r_{_\mathrm H}^2}{3})}{r_{_\mathrm H}^2+a^2}\,,
\end{align} 
in which $r_{_{\mathrm{H}}}$ is the radius of the considered horizon. The chosen horizon can be inner or outer event or cosmological horizon. By invoking the Hawking temperature(s) $T_{_\mathrm{H}}=\frac{\kappa_{_\mathrm{H}}}{2\pi}$ (here $\mathrm{H}$ in $T_{_\mathrm{H}}$ refers to the ``Horizon") \cite{Gibbons:1977mu}, these BHs satisfy the first law(s) of thermodynamics
\begin{equation*}
 \delta M=T_{_\mathrm{H}}\delta S_{_\mathrm H}+\Omega_{_\mathrm H}\delta J\,,
 \end{equation*}
for each one of the horizons.  
In the case of $\Lambda<0$, $\Lambda=0$, and $\Lambda>0$ and for legitimate parameters $m$ and $a$, geometry and thermodynamics of the Kerr-AdS, Kerr, and Kerr-dS BHs are found respectively.

The paper is organized as follows. In Sec. \ref{sec-review}, a quick but practical review of SPSM is presented. In Sec. \secref{sec-K-dS}, thermodynamics of Kerr-dS BHs is analyzed. In Sec. \ref{sec-K-N-dS}, the analysis is extended to the electrically charged BHs. Finally, we conclude the analysis in Sec. \ref{sec-conclude}. A more conceptual review of SPSM is provided in Appendix \ref{app-deep}.  

\section{A practical review on solution phase space method}\label{sec-review}
In this section, we provide a review on the ``solution phase space method" which is a recent method for calculating conserved charges associated with ``exact symmetries," proposed in Ref. \cite{HS:2015xlp} (and its precursor Ref. \cite{Hajian:2015eha}). For the ease of the reader, the review is presented in this way: here, the method is described as a very quick and shallow but practical instruction followed by a simple example, in order to illustrate simplicity and generality of the method. In Appendix \ref{app-deep}, we will provide a more conceptual review of the method. For the full coverage of the concepts in the formulation, it is recommended to refer to the original work \cite{HS:2015xlp}.  

Consider a generally covariant gravitational theory in $d$-dimensional spacetime, with some probable internal gauge symmetries, \emph{e.g.} some Maxwell-like gauge fields $A^a$ labelled by the index $a$. The dynamical fields (\emph{e.g.} the metric, gauge fields, scalar fields, \dots) can be denoted collectively by $\Phi(x^\mu)$. In addition, consider a family of BH solutions to this theory, represented by the dynamical fields $\hat\Phi(x^\mu;p_j)$. As the notation suggests, we assume that dynamical fields are identified by some parameters $p_j$  up to unphysical coordinate transformations.  \emph{Exact symmetries} for such a family of solutions would be transformations generated by $\eta$ such that
\begin{equation}\label{exact symmetry}
\eta=\{\zeta,\lambda^a\}\,, \qquad \delta_\eta\hat\Phi\equiv \mathscr{L}_\zeta\hat\Phi+\delta_{\lambda^a}\hat A^a=0
\end{equation}
in which $\zeta=\zeta^\mu\partial_\mu$ is a vector field over spacetime and $\lambda^a$ are some scalars generating gauge transformations $A^a\to A^a+\mathrm{d} \lambda^a$. Hence, exact symmetries are those diffeomorphism+gauge transformations whose combination do not change the dynamical fields at all.

Focusing on a BH $\hat\Phi(x^\mu;p_j)$ in the family mentioned above as a solution to the theory under consideration, one can attribute variations of a conserved charge to each one of the exact symmetry generators $\eta$, denoted by $\hat\delta H_\eta$\footnote{The choice of the alphabet $H$ originates from the words ``Hamiltonian generator" which we use interchangeably with the ``conserved charge".}. To this end, one can follow the instructions below.
\begin{enumerate}
\item One needs to calculate a $(d\!-\!2)$-form associated with the theory, and for a generic diffeomorphism+gauge transformation $\epsilon=\{\xi,\lambda^a\}$ (not necessarily an exact symmetry generator) \cite{Lee:1990gr,Iyer:1994ys,Wald:1999wa,Barnich:2001jy}. This $(d\!-\!2)$-form is denoted by $\boldsymbol{k}_\epsilon(\delta\Phi,\Phi)$. $\Phi$ and $\delta \Phi$ have to satisfy e.o.m and linearized e.o.m respectively. The $\boldsymbol{k}_\epsilon$ is pragmatically the most important form which one needs to calculate the conserved charges. Calculation of the $\boldsymbol{k}_\epsilon(\delta\Phi,\Phi)$ is a standard subject.  In this practical review, we will not delve into its conceptual origin and explicit definition. We postpone this important issue to the Appendix \ref{app-deep}, which its major part is devoted to the introduction of $\boldsymbol{k}_\epsilon$ and its explicit calculation for the simple Einstein-Hilbert theory, which is the relevant theory for our discussions. Therefore, here, we assume that it is known for the theory under consideration. The nice thing is that once $\boldsymbol{k}_\epsilon$  is calculated, it can be used for any solution to the theory. 
\item For the chosen solution, the ``parametric variations" are needed to be calculated, which are denoted by $\hat\delta \Phi$. They can be calculated simply by  differentiating dynamical fields with respect to the parameters \cite{HSS:2014twa},
\begin{equation}\label{parametric variations}
\hat{\delta}\Phi\equiv \frac{\partial {\hat\Phi}}{\partial p_j}\delta p_j.
\end{equation}
\item  The dynamical fields $\Phi$, the perturbations $\delta \Phi$, and the generator $\epsilon$ in the $\boldsymbol{k}_\epsilon(\delta\Phi,\Phi)$ should be replaced by the black hole solution $\hat\Phi$,  parametric variations $\hat \delta \Phi$, and exact symmetry generator $\eta$  respectively. Then, an integration over an arbitrary closed, smooth, and spacelike $d-2$-dim surface $\partial \Sigma$ surrounding the BH singularity should be taken. Mathematically,
\begin{equation}\label{hat delta H}
 \hat\delta H_{\eta}= \oint_{\partial \Sigma}\boldsymbol{k}_\eta(\hat\delta\Phi,\hat\Phi)\,.
\end{equation}
\item Integration above would be unambiguous, and a function of parameters $p_j$. It would  not be a function of the spacetime coordinates (the reason is postponed to Appendix \ref{app-deep}), \emph{i.e.} $\hat\delta H_{\eta}=\hat\delta H_{\eta}(p_j)$. If this function is integrable over the parameters, then the conserved charge $H_\eta$ would be found by an integration over the parameters, \emph{i.e.}
\begin{equation}\label{finite H xi}
H_\eta[\hat\Phi(p)]=\int_{\bar p}^{p} \hat\delta H_\eta+H_\eta[\bar\Phi(\bar p)]\,,
\end{equation}
The $H_\eta[\bar\Phi]$ is the reference point (\emph{i.e.} constant of integration) for the $H_\eta$ defined on some specific reference field configuration $\bar\Phi(x^\mu;\bar p_j)$.   
\end{enumerate}

For clarity let us give a simple example. Einstein-Hilbert (EH) theory is a generally covariant  gravitational theory which is described by the Lagrangian $\mathcal{L}=\frac{1}{16\pi G} R$ without any gauge fields present.  The Kerr BH, which can be represented by the metric
\begin{align}\label{Kerr metric}
\hspace{-0.23cm}&\mathrm{d}s^2= -(1\!-\! f)\mathrm{d}t^2+\frac{\rho ^2}{\Delta_r}\mathrm{d}r^2+{\rho ^2} \mathrm{d}\theta ^2 -2 fa\sin ^2 \theta\,\mathrm{d}t \mathrm{d}\varphi+\left(r^2+a^2+fa^2\sin ^2\theta \right)\sin ^2\theta\,\mathrm{d}\varphi ^2\,,\nonumber\\
&\hspace*{1.7cm}\rho^2 \equiv r^2+a^2 \cos^2 \theta\,,\qquad \Delta_r \equiv r^2+a^2-2Gmr\,,\qquad
f\equiv\frac{2Gmr}{\rho ^2}\,,
\end{align}
is a solution to the equation of motion of this theory in $4$-dimensions \cite{Kerr:1963ud}, and is identified by two parameters $p_{_1}\equiv m$ and $p_{_2}\equiv a$. So, the dynamical field would be the metric $\hat g_{\alpha\beta}(x^\mu; m,a)$. The horizons are at the radii $r_{\pm}=Gm\pm\sqrt{G^2m^2-a^2}$. Exact symmetries for this solution are the stationarity and axial isometry (generated by the Killing vectors $\zeta=\partial_t$ and $\zeta=\partial_\varphi$) and any linear combination of them. 
Let us use the instructions above to find the mass $M$ and angular momentum $J$ for our simple BH; the Kerr BH. The first step: the theory is the EH theory. There is not any gauge field present, so $\epsilon$ would be $\{\xi,0\}$ for some arbitrary vector field  $\xi=\xi^\mu\partial_\mu$. Hence for bookkeeping, we can use $\xi$ instead of $\epsilon=\{\xi,0\}$. The $(d\!-\!2)$-form $\boldsymbol{k}_\xi(\delta g_{\alpha\beta},g_{\alpha\beta})$ is well-known for the EH theory to be
\begin{equation}\label{k xi EH}
\boldsymbol{k}^\text{EH}_\xi(\delta g_{\alpha\beta},g_{\alpha\beta})=\frac{\sqrt{-g}}{2!\,2!}\,\,\epsilon_{\mu\nu\sigma\rho} \,k_\xi^{\text{EH}\,\mu\nu}\,\,\mathrm{d}x^{\sigma}\wedge \mathrm{d}x^{\rho} 
\end{equation}
where
\begin{align}
&k_\xi^{\text{EH}\,\mu\nu}(\delta g_{\alpha\beta},g_{\alpha\beta})=\dfrac{1}{16 \pi G}\Big(\Big[\xi^\nu\nabla^\mu h
-\xi^\nu\nabla_\tau h^{\mu\tau}+\xi_\tau\nabla^{\nu}h^{\mu\tau}+\frac{1}{2}h\nabla^{\nu} \xi^{\mu}-h^{\tau\nu}\nabla_\tau\xi^{\mu}\Big]-[\mu\leftrightarrow\nu]\Big)\label{k EH integrand}
\end{align} 
in which $h^{\mu\nu}\equiv g^{\mu \sigma}g^{\nu\tau}\delta g_{\sigma \tau}$ and $h\equiv h^\mu_{\,\,\mu}$. The $\epsilon_{\mu\nu\sigma\rho}$ is the Levi-Civita symbol in Kerr geometry, \emph{i.e.} $\epsilon_{tr\theta\varphi}=+1$ and changes sign under odd permutations of the indices.  Nonetheless, for completeness of calculations, we have provided the step-by-step and detailed derivation of $\boldsymbol{k}^\text{EH}_\xi$ in Appendix \ref{app-deep}. The second step in the instruction is very simple: For the Kerr solution, parametric variations are
\begin{equation}\label{Kerr parametric var}
\hat\delta g_{\alpha\beta}=\frac{\partial \hat g_{\alpha\beta}}{\partial m}\delta m+\frac{\partial \hat g_{\alpha\beta}}{\partial a}\delta a\,,
\end{equation}
in which $\hat g_{\alpha\beta}$ is the Kerr metric \eqref{Kerr metric}. The third step starts by the replacement $g_{\alpha\beta}\to \hat g_{\alpha\beta}$, $\delta g_{\alpha\beta}\to \hat \delta g_{\alpha\beta}$ and $\xi\to \zeta$ in the $\boldsymbol{k}^\text{EH}_\xi$ reported in the Eq. \eqref{k xi EH}, for some arbitrary Killing vector field $\zeta$. Then, the integration over a closed smooth $2$-dim surface $\partial \Sigma$ surrounding the singularity of the BH, \emph{i.e.} enclosing the $r=0$, should be calculated. For simplicity of calculations, we can take the $\partial \Sigma$ to be any surface of constant $(t,r)$ for $r>0$. Therefore, the final expression to be calculated would be
\begin{align}\label{Kerr charge calc}
\hat\delta H_{\zeta}= \oint_{\partial \Sigma}\boldsymbol{k}^\text{EH}_\zeta(\hat\delta g_{\alpha\beta},\hat g_{\alpha\beta}) =\int_0^{2\pi}\int_{0}^{\pi}  \sqrt{-g}\,k_\zeta^{\text{EH}\,tr}(\hat\delta g_{\alpha\beta},\hat g_{\alpha\beta})\,\mathrm{d}\theta \,\mathrm{d}\varphi\,,
\end{align}
in which $k_\zeta^{\text{EH}\,tr}$ is simply the $tr$ component of the Eq. \eqref{k EH integrand}. Thanks to the linearity of the $\hat\delta H_{\zeta}$ in terms of $\hat\delta g_{\alpha\beta}$, the parametric variations \eqref{Kerr parametric var} can be inserted into the Eq. \eqref{Kerr charge calc} {term by term}, simplifying the calculations. The fourth step would be integrating the result over the parameters $m$ and $a$. The results are reported below.
\paragraph{Mass and angular momentum:} Choosing the Killing vectors $\eta_{_M}=\partial_t$ and  $\eta_{_J}=\partial_\varphi$, the $\hat \delta M$ and $\hat{\delta}J$ are found respectively
\begin{align}
 \hat\delta M\equiv \hat\delta H_{\eta_{_M}}=1\times \delta m+0\times \delta a \qquad &\Rightarrow\qquad  M=m\,,\\
 \hat\delta J\equiv -\hat\delta H_{\eta_{_J}}=a\times \delta m+m\times \delta a \qquad &\Rightarrow\qquad  J=ma\,.
\end{align}
The relative minus sign in the definition of the angular momentum is the standard unimportant convention. The reference point of the mass, $H_{\eta_{_M}}[\bar g_{\alpha\beta}]$, has been chosen to vanish for solution $\bar g_{\alpha\beta}$ identified by $m=a=0$, \emph{i.e.} the Minkowski spacetime. Also, the reference point of the angular momentum, $H_{\eta_{_J}}[\bar g_{\alpha\beta}]$, has been chosen to vanish for the Minkowski spacetime.
\paragraph{Entropies:} Let us denote the $r_{\pm}$ collectively as $r_{_\mathrm{H}}$. In SPSM, in the absence of gauge symmetries, entropy is defined as the Hamiltonian generator associated with the normalized horizon Killing vector $\eta_{_\mathrm{H}}=\frac{\zeta_{_\mathrm{H}}}{\kappa_{_\mathrm{H}}}$ \cite{HS:2015xlp,Wald:1993nt,Iyer:1994ys}. Hence, for each one of the horizons in the Kerr solution one can calculate an entropy. For the Kerr solution, surface gravity, angular velocity and the $\zeta_{_\mathrm{H}}$ are explicitly 
\begin{equation}\label{Kerr zeta H}
 \kappa_{_\mathrm{H}}=\frac{r_{_\mathrm{H}}^2-a^2}{2r_{_\mathrm{H}}(r_{_\mathrm{H}}^2+a^2)}\,,\qquad \Omega_{_\mathrm{H}}=\frac{a}{r_{_\mathrm{H}}^2+a^2}\,,\qquad  \zeta_{_\mathrm{H}}=\partial_t+\Omega_{_\mathrm{H}}\partial_\varphi\,. 
\end{equation}
The same procedure as the mass and angular momentum for the $\eta_{_\mathrm{H}}$ yields
 \begin{align}
 \hat\delta S_{_\mathrm{H}}\equiv \hat\delta H_{\eta_{_\mathrm{H}}}=\frac{\partial\Big(\frac{\pi(r_{_\mathrm{H}}^2+a^2)}{G}\Big)}{\partial m}\delta m+\frac{\partial\Big(\frac{\pi(r_{_\mathrm{H}}^2+a^2)}{G}\Big)}{\partial a}\delta a \qquad \Rightarrow\qquad  S_{_\mathrm{H}}=\frac{\pi(r_{_\mathrm{H}}^2+a^2)}{G}\,.
 \end{align} 
In the last step, integration over parameters has been calculated such that reference points would vanish for the Minkowski spacetime.  
\paragraph{First law(s):} Using the relation between generators of the entropy, mass, and angular momentum, \emph{i.e.} 
\begin{equation}
\eta_{_\mathrm{H}}=\frac{2\pi}{\kappa_{_\mathrm{H}}}(\eta_{_M}+\Omega_{_\mathrm{H}}\eta_{_J})\,,
\end{equation}
 and noticing the linearity of $\delta H_\xi$ in $\xi$ (see Eq. \eqref{k xi EH}), the first law associated with the horizon at $r_{_\mathrm{H}}$ would simply follow as 
\begin{equation}\label{Kerr first law}
\delta M=T_{_\mathrm{H}}\delta S_{_\mathrm{H}}+\Omega_{_\mathrm{H}}\delta J\,, 
\end{equation} 
in which the Hawking temperature(s) $T_{_\mathrm{H}}=\frac{\kappa_{_\mathrm{H}}}{2\pi}$ are used. Notice that depending on the choice of inner or outer horizon, there are two versions of the first law. Nonetheless, it is usual to identify the equation corresponding to the outer (event) horizon as 
the first law of thermodynamics, due to the positivity of the temperature. Also note that the only condition on the perturbations $\delta g_{\alpha\beta}$ in Eq. \eqref{Kerr first law} is satisfying linearized e.o.m. So, they include parametric variations $\hat \delta g_{\alpha\beta}$, in addition to other dynamically allowed perturbations.

We encourage the reader to repeat the steps above for the Kerr-AdS BHs, in order to realize simplicity and reliability of the formulation. A good news is that $\boldsymbol{k}_\epsilon(\delta\Phi,\Phi)$ in Eq. \eqref{k xi EH} would be independent of the choice of cosmological constant $\Lambda$ in the EH theory. Hence, the first step in the instructions is already done.   

Solution Phase Space Method works similarly for many other (not necessarily BH) solutions to generally covariant gravitational theories, in any dimensions and with generic asymptotic behaviors. To see more examples including BTZ, Kerr-Newman, Kerr-AdS, $5$-dim Myers-Perry, $4$-dim Kaluza-Klein BHs and near horizon geometry of their extremal cases, the papers \cite{HS:2015xlp,Hajian:2015eha} can be referred.  A conceptual review of SPSM is provided in Appendix \ref{app-deep}. In the next section, we use this method to study the thermodynamics of the Kerr-dS BHs. 

\section{Conserved charges and first law(s) for Kerr-dS black holes}\label{sec-K-dS}
Equipped with the SPSM, we can find the Hamiltonian generators labelling the Kerr-dS BHs, \emph{i.e.} mass $M$ and angular momentum $J$. Besides, using the normalized horizon Killing vectors $\eta_{_\mathrm{H}}$, the entropies $S_{_\mathrm{H}}$ can be found. For sure, to find the finite integrated results, variations of the mentioned conserved charges have to be integrable. Finally, the first law(s) of thermodynamics as identities relating variations of these charges will be presented, although their general proof in the context of SPSM is very simple and has been provided in Ref. \cite{HS:2015xlp}.

Kerr-dS BH is a $4$-dimensional solution to the gravitational theory described by the Lagrangian density $\mathcal{L}=\frac{1}{16\pi G} (R-2\Lambda)$, in which $R$ and $\Lambda>0$ are Ricci scalar and cosmological constant respectively \cite{Griffiths:2009ex}. This BH has only one dynamical field, which is the metric 
\begin{align}\label{Kerr-dS metric}
\hspace{-0.23cm}\mathrm{d}s^2= -&\Delta_\theta(\frac{1-\frac{r^2}{l^2}}{\Xi}\!-\!\Delta_\theta f)\mathrm{d}t^2+\frac{\rho ^2}{\Delta_r}\mathrm{d}r^2+\frac{\rho ^2}{\Delta_\theta} \mathrm{d}\theta ^2 -2\Delta_\theta fa\sin ^2 \theta\,\mathrm{d}t \mathrm{d}\varphi\nonumber \\
+&\left( \frac{r^2+a^2}{\Xi}+fa^2\sin ^2\theta \right)\sin ^2\theta\,\mathrm{d}\varphi ^2\,,
\end{align}
where
\begin{align}
\rho^2 &\equiv r^2+a^2 \cos^2 \theta\,,\qquad \Delta_r \equiv (r^2+a^2)(1-\frac{r^2}{l^2})-2Gmr\,,\nonumber\\
\Delta_\theta&\equiv 1+\frac{a^2}{l^2}\cos ^2\theta\,,\qquad
 \Xi\equiv 1+\frac{a^2}{l^2}\,,\qquad f\equiv\frac{2Gmr}{\rho ^2\Xi^2}\,.\nonumber
\end{align}
This metric has two parameters $p_{_1}=m$ and $p_{_2}=a$. They are free parameters up to some physical constraints \cite{Booth:1998gf}, which are unimportant in our discussion. Radius of the dS$_4$ has been denoted by $l$, which is related to the $\Lambda$ by the relation $\Lambda=\frac{3}{l^2}$.  The metric is written in coordinates such that the BH be nonrotating with respect to infinity, \emph{i.e.} at the $r\to \infty$. As we will see in a moment, these coordinates have the nice property that exact symmetries to which mass and angular momentum are associated have the simple form of $\partial_t$ and $\partial_\varphi$ respectively. Putting this issue aside, according to the covariance of the SPSM, the choice of coordinates would be irrelevant to the calculation of conserved charges.  
  
The SPSM instructions for calculating the  charges can be performed easily as below.
\begin{enumerate}
\item For the EH gravity with arbitrary cosmological constant, the $\boldsymbol{k}_\epsilon(\delta\Phi,\Phi)$ is exactly the Eq. \eqref{k xi EH} (see Appendix \ref{app-deep}). 
\item The parametric variations can be simply found by Eq. \eqref{Kerr parametric var} in which $\hat g_{\alpha\beta}$ would be the Kerr-dS metric Eq. \eqref{Kerr-dS metric}.
\item For the specific choices of the exact symmetries $\eta_{_\mathrm{M}}=\partial_t$ and $\eta_{_J}=\partial_\varphi$, mass and angular momentum variations can be found by Eq. \eqref{Kerr charge calc} to be
\begin{align}
 \hat\delta M\equiv \hat\delta H_{\eta_{_M}}=\frac{\partial \big(\frac{m}{\Xi^2}\big)}{\partial m}\delta m+\frac{\partial \big(\frac{m}{\Xi^2}\big)}{\partial a}\delta a=\hat \delta(\frac{m}{\Xi^2}) \,,\label{Kerr-dS mass var}\\
 \hat\delta J\equiv -\hat\delta H_{\eta_{_J}}=\frac{\partial \big(\frac{ma}{\Xi^2}\big)}{\partial m}\delta m+\frac{\partial \big(\frac{ma}{\Xi^2}\big)}{\partial a}\delta a=\hat \delta (\frac{ma}{\Xi^2}) \,.\label{Kerr-dS angular var}
\end{align}
\item By integration over parameters (which is basically integration over solution phase space \cite{HS:2015xlp}), finite results are found to be unambiguously
\begin{equation}
M=\frac{m}{\Xi^2}\,,\qquad \quad J=\frac{ma}{\Xi^2}\,.
\end{equation}
The reference fields (constant of integrations) are chosen to be vanishing mass and angular momentum for $m=a=0$, \emph{i.e.} the pure dS$_4$ spacetime.
\end{enumerate}
Notice that the results, which are reported above, are independent of the chosen surface of integration $\partial \Sigma$.  Hence, the $M$ and $J$ can be considered as charges attributed to the geometry as a whole, irrespective to any specific horizon or asymptotics. Moreover, assuming $0\!\leq\!m$ and $0\leq a$, mass and angular momentum are positive (\emph{cf.} negative results in the literature). 

Calculation of the entropy would be similar to the mass and angular momentum, but for the normalized horizon Killing vector $\eta_{_\mathrm{H}}=\frac{2\pi}{\kappa_{_\mathrm{H}}}\{\zeta_{_\mathrm{H}},0\}$ in which
\begin{equation}\label{Kerr-dS zeta}
 \kappa_{_\mathrm{H}}=\frac{r_{_\mathrm H}(1-\frac{a^2}{l^2}-3\frac{r_{_\mathrm H}^2}{l^2}-\frac{a^2}{r_{_\mathrm H}^2})}{2(r_{_\mathrm H}^2+a^2)}\,,\qquad \Omega_{_\mathrm H}=\frac{a(1-\frac{r_{_\mathrm H}^2}{l^2})}{r_{_\mathrm H}^2+a^2}\,, \qquad \zeta_{_\mathrm H}=\partial_t+\Omega_{_\mathrm H}\partial_\varphi\,, 
\end{equation}
where $\kappa_{_\mathrm{H}}$ and $\Omega_{_\mathrm{H}}$ are surface gravity and angular velocity on the chosen horizon respectively \cite{Gibbons:1977mu}. In the above, $r_{_\mathrm H}$ is radius of the chosen horizon as a solution to $\Delta_r=0$, explicitly
\begin{equation}\label{r-H-Kerr-dS}
(r_{_\mathrm{H}}^2+a^2)(l^2-r_{_\mathrm{H}}^2)-2Gm\ell^2 r_{_\mathrm{H}}=0.
\end{equation}
In other words, $r_{_\mathrm H}$ can be the radius of \emph{any} one of the horizons present in the Kerr-dS geometry, for the specific choice of $m$ and $a$ (see \cite{Booth:1998gf} for detailed analysis).  We will discuss more on this issue at the end of this section. The result of the calculations turns out to be
\begin{equation}\label{Kerr dS del S}
\hat\delta S_{_\mathrm{H}}\equiv \hat \delta H_{\eta_{_\mathrm{H}}}=\frac{\partial \left(\frac{\pi(r_{_\mathrm{H}}^2+a^2)}{G\,\Xi}\right)}{\partial m}\delta m + \frac{\partial \left(\frac{\pi(r_{_\mathrm{H}}^2+a^2)}{G\,\Xi}\right)}{\partial a}\delta a=\hat \delta \left(\frac{\pi(r_{_\mathrm{H}}^2+a^2)}{G\,\Xi}\right)\,.
\end{equation}
 Therefore, one may integrate $\hat \delta H_{\eta_{_\mathrm{H}}}$ over parameters/solution phase space, to obtain the corresponding charge. For any choice of horizon, the reference point of the entropy can be chosen such that
\begin{equation}\label{Kerr dS entropy}
S_{_\mathrm{H}}\equiv H_{\eta_{_\mathrm{H}}}=\frac{\pi(r_{_\mathrm{H}}^2+a^2)}{G\,\Xi}\,.
\end{equation}
 {For example, in the case of \emph{event} horizon of the BH,  by choosing the pure dS$_4$ spacetime as the reference point with vanishing entropy, \emph{i.e.} $H_{\eta_{_\mathrm{H}}}[\text{dS}_4]=0$}, integrating Eq. \eqref{Kerr dS del S} results in the standard entropy for Kerr-dS BH \eqref{Kerr dS entropy}. On the other hand, for the case of cosmological horizon, the reference point can be chosen to be $H_{\eta_{_\mathrm{H}}}[\text{dS}_4]=\frac{\pi l^2}{G}$ in order to reproduce Eq. \eqref{Kerr dS entropy}, which is the well-known Gibbons-Hawking entropy of cosmological event horizon \cite{Gibbons:1977mu}.

In the context of SPSM, the first law(s) of BH thermodynamics enjoy a very simple proof \cite{HS:2015xlp}. It is so simple that it can be explained in words, in one sentence! It is: $\delta H_{\xi}$ is linear in its generator vector field $\xi$, so the relation $\eta_{_\mathrm H}=\frac{2\pi}{\kappa_{_\mathrm{H}}}(\eta_{_M}+\Omega_{_\mathrm H}\eta_{_J})$ directly leads to $\delta S_{_\mathrm{H}}\!=\!\frac{2\pi}{\kappa_{_\mathrm{H}}}\delta M\!-\!\frac{2\pi}{\kappa_{_\mathrm{H}}}\Omega_{_\mathrm H}\delta J$, which by the Hawking temperature(s) $T_{_\mathrm{H}}\!=\!\frac{\kappa_{_\mathrm{H}}}{2\pi}$ yields the first law(s)
\begin{equation}\label{first law KdS}
 \delta M=T_{_\mathrm{H}}\delta S_{_\mathrm{H}}+\Omega_{_\mathrm H}\delta J\,.
 \end{equation} 
Notice that this proof works for any perturbation which satisfies the linearized equation of motion, including parametric variations, but is not limited to them. To cross check, one can investigate Eq.\eqref{first law KdS} for the parametric variations, using Eq. \eqref{Kerr-dS mass var}, Eq. \eqref{Kerr-dS angular var}, and Eq. \eqref{Kerr dS del S}.

If  the reader has already followed the calculations for the Kerr-AdS BHs (which can be found in Refs. \cite{HS:2015xlp,Hajian:2015eha}), she/he might have found that the analysis for the Kerr-dS BHs is exactly similar, if one keeps an abstract and undetermined $\Lambda$ in calculations. It is basically because the SPSM is insensitive to the chosen $\Lambda$. As a result, as far as conserved charges and the first law(s) of thermodynamics are concerned, we can unify the $4$-dimensional stationary BHs with either AdS, flat or dS asymptotics. The unified results has been presented in Sec. \ref{sec Intro} and we will not repeat here.

Here, it is a good place to discuss about two questions. The first question is about the choice of $\partial_t$ as the exact symmetry generator of the mass. In the asymptotic de Sitter BH geometries, there is not a clear choice for this generator, in contrast to \emph{e.g.} asymptotic flat cases. The problem originates from the different signature of the metric outside the cosmological horizon. Nonetheless, SPSM enables us to choose the correct generator. Specifically, integrability condition imposes strong constraint on the choice of the generator, which rules out other proposed candidates (see Ref. \cite{Chrusciel:2015sna} and Refs. therein). For example, if one uses the time translation in the Boyer-Lindquist coordinates (\emph{e.g.} see Refs. \cite{Ghezelbash:2004af,Akcay:2010vt} to find explicit metric in this coordinates) as the generator for the mass, then the result would not be integrable.  To see the reason, denoting the time in Boyer-Lindquist coordinates by $\tau$, it can be found that $\partial_{\tau}=\partial_t-\frac{a}{l^2} \partial_\varphi$. Then, by linearity of $\delta H_\epsilon$ in $\epsilon$
\begin{equation}
\hat \delta H_{\partial_\tau}=\hat \delta H_{\partial_t}-\frac{a}{l^2}\hat \delta H_{\partial_\varphi}\,.
\end{equation}
Replacing the right hand side from Eq. \eqref{Kerr-dS mass var} and \eqref{Kerr-dS angular var}, the result would be manifestly non-integrable on parameters. We will refer the reader, who is interested in rigorous integrability calculations, to the  Eq. \eqref{integrability cond modified} in Appendix \ref{app-deep}, and also to the main reference \cite{HS:2015xlp}. It is also worth mentioning that in addition to the integrability, another guide for the choice of the correct generator with the correct sign would be paying attention to the unified description of thermodynamics of Kerr-AdS, Kerr, and Kerr-dS BHs; in Kerr and Kerr-AdS we know how to fix the mass generator. Hence by smooth change of $\Lambda$, we can find the appropriate generator in the Kerr-dS BH. 

The second question is about how we can choose the horizon responsible for defining the ``usual entropy" and ``usual first law" for the Kerr-dS BH. As it was emphasized earlier, in SPSM, one can associate ``entropy" variations to any one of the horizons present in the geometry, including BH or cosmological horizons, utilizing associated normalized horizon Killing vector $\eta_{_\mathrm{H}}$. This interesting freedom is also present in the choice between outer and inner horizons in the usual BH geometries without cosmological horizon, \emph{e.g.} the Kerr and Kerr-AdS BHs. Moreover, an identity analogous to the first law  would also follow for each one of the horizons, simply because of linearity of charge variations in their generators, accompanied by the freedom of the integration surface for all charges including the entropies. After that, in the case of integrable entropy variations (which is generically true for any horizon), the finite entropy can be found.  Although these first laws are not independent identities, but the choice of event horizon, \emph{i.e.} the outermost horizon of BHs, is an standard choice in order to have positive surface gravity, and so positive temperature. For example, this is the choice for the Kerr and Kerr-AdS BHs. We can request similar choice for the Kerr-dS BH for the same reason. Besides, one might reach to the same choice  by requesting temperature and entropy to be  \emph{continuous} functions of $\Lambda$ when $\Lambda$ changes sign. So the choice of event horizon for Kerr-AdS and Kerr BHs would be extended to the similar choice for Kerr-dS BH. Nevertheless, it can be an interesting line of research to understand physical implications of different entropies associated with different horizons, and their relation to the microstates of the system.

Generalization of the analysis presented in this section to $4<d$ dimensional spacetime is straightforward, because this generalization for calculation of charges in SPSM is straightforward. Hence, one would not expect anything new in that analysis, and the unification can be taken as granted although it needs direct check. In this paper, we will not ensue this line of generalization. Instead, we generalize the analysis to include electric charge in the next section.

\section{Conserved charges and first law(s) for Kerr-Newman-dS black holes}\label{sec-K-N-dS}
Here, we analyse Kerr-Newman BHs with AdS, flat or dS asymptotics in the unified picture achieved in the preceding section. We will highlight the main points and results. So, repeated technical details of calculations might be ignored.  The theory under consideration would be the Einstein-Maxwell-$\Lambda$ theory, which is described by the Lagrangian density $\mathcal{L}=\frac{1}{16\pi G} (R-F^2-2\Lambda)$. The $F=\mrd A$ is the electromagnetic field strength, and $A=A_\mu \mrd x^\mu$ is the gauge field. Dynamical fields of the mentioned  BH solutions are the metric $\hat g_{\alpha\beta}(x^\mu;m,a,q)$ and the gauge field $\hat A_\alpha(x^\mu;m,a,q)$, collectively denoted by $\hat \Phi$. The metric in asymptotically-nonrotating coordinates would be similar to the metric \eqref{general metric} by the replacement $\Delta_r \equiv (r^2+a^2)(1-\frac{\Lambda r^2}{3})-2Gmr+q^2$ \cite{Carter:1970ea}. In these coordinates, the gauge field would be 
\begin{equation}
\hat A_\mu \mrd x^\mu=\frac{qr}{\rho^2\Xi}(\Delta_\theta\mrd t-a\sin^2 \theta \,\mrd \varphi)\,. 
\end{equation} 
These dynamical fields satisfy the equations of motion
\begin{align}
R_{\mu\nu}-\frac{1}{2}(R-2\Lambda)g_{\mu\nu}\!=2F_{\mu\alpha}F^{\,\,\alpha}_{\nu}-\frac{1}{2}F^2 g_{\mu\nu}, \qquad\qquad  \nabla_\alpha F^{\alpha\mu}=0\,.
\end{align}
For the theory under consideration, and for diffeomorphism+gauge transformation $\epsilon=\{\xi,\lambda\}$ 
\begin{equation}
\boldsymbol{k}_\epsilon(\delta \Phi,\Phi)=\frac{\sqrt{-g}}{2!\,2!}\,\,\epsilon_{\mu\nu\sigma\rho} \,(k_\epsilon^{\text{EH}\,\mu\nu}+k_\epsilon^{\text{M}\,\mu\nu})\,\,\mathrm{d}x^{\sigma}\wedge \mathrm{d}x^{\rho} 
\end{equation}
where $k_\epsilon^{\text{EH}\,\mu\nu}$ is the one in Eq. \eqref{k EH integrand}, and
\begin{equation}
k_\epsilon^{\text{M}\,\mu\nu}=\frac{1}{8 \pi G}\Big(\Big[\big(\frac{-h}{2} F^{\mu\nu}+2F^{\mu\rho}h_\rho^{\;\;\nu}-\delta F^{\mu\nu}\big)({\xi}^\sigma A_\sigma+\lambda)- F^{\mu\nu}\xi^\rho \delta A_\rho-2F^{\rho\mu}\xi^\nu \delta A_\rho\Big]-[\mu\leftrightarrow\nu]\Big)\,.
\end{equation}
Notice that in the equation above
\begin{equation}
\delta F^{\mu\nu}\equiv g^{\mu\alpha}g^{\nu\beta}\delta F_{\alpha\beta}=g^{\mu\alpha}g^{\nu\beta}(\delta \mrd A)_{\alpha\beta}=g^{\mu\alpha}g^{\nu\beta}(\mrd \delta A)_{\alpha\beta}\,.
\end{equation}
The solutions which we have focused on are identified by three parameters $(m,a,q)$. So, parametric variations $\hat \delta \Phi$ are
\begin{equation}\label{K-N-dS parametric var}
\hat\delta g_{\alpha\beta}=\frac{\partial \hat g_{\alpha\beta}}{\partial m}\delta m+\frac{\partial \hat g_{\alpha\beta}}{\partial a}\delta a+\frac{\partial \hat g_{\alpha\beta}}{\partial q}\delta q\,, \qquad \hat\delta A_{\mu}=\frac{\partial \hat A_{\mu}}{\partial m}\delta m+\frac{\partial \hat A_{\mu}}{\partial a}\delta a+\frac{\partial \hat A_{\mu}}{\partial q}\delta q\,.
\end{equation}
\paragraph{Mass and angular momentum:} Calculating Hamiltonian generators for exact symmetries generated by $\eta_{_M}=\{\partial_t,0\}$ and  $\eta_{_J}=\{\partial_\varphi,0\}$ on any $2$-dimensional spacelike smooth surface $\partial \Sigma$, the $\hat \delta M$ and $\hat{\delta}J$ are unambiguously found, which are respectively as
\begin{align}
 \hat\delta M=\frac{\partial \big(\frac{m}{\Xi^2}\big)}{\partial m}\delta m+\frac{\partial \big(\frac{m}{\Xi^2}\big)}{\partial a}\delta a +\frac{\partial \big(\frac{m}{\Xi^2}\big)}{\partial q}\delta q =\hat \delta \big(\frac{m}{\Xi^2}\big)\qquad &\Rightarrow\qquad  M=\frac{m}{\Xi^2}\,,\\
 \hat\delta J=\frac{\partial \big(\frac{ma}{\Xi^2}\big)}{\partial m}\delta m+\frac{\partial \big(\frac{ma}{\Xi^2}\big)}{\partial a}\delta a +\frac{\partial \big(\frac{ma}{\Xi^2}\big)}{\partial q}\delta q =\hat \delta \big(\frac{ma}{\Xi^2}\big)\qquad &\Rightarrow\qquad  J=\frac{ma}{\Xi^2}\,.
\end{align}
The reference points have been chosen such that pure dS$_4$ spacetime would have vanishing mass and angular momentum. 
\paragraph{Electric charge:} By the choice of $\eta_{_Q}=\{0,1\}$, \emph{i.e.} the global part of the gauge transformations, and integrating on $\partial \Sigma$ mentioned above, electric charge is found as
\begin{equation}
\hat\delta Q\equiv \hat \delta H_{\eta_{_Q}}=\frac{\partial \big(\frac{q}{\Xi}\big)}{\partial m}\delta m+\frac{\partial \big(\frac{q}{\Xi}\big)}{\partial a}\delta a +\frac{\partial \big(\frac{q}{\Xi}\big)}{\partial q}\delta q =\hat \delta \big(\frac{q}{\Xi}\big) \qquad \Rightarrow\qquad Q=\frac{q}{\Xi}\,.
\end{equation}
The reference points is clearly vanishing electric charge for the pure dS$_4$ spacetime.
\paragraph{Entropies:} The horizon surface gravities, angular velocities and electric potentials are respectively 
\begin{equation}\label{Kerr-N-dS zeta}
 \kappa_{_\mathrm{H}}=\frac{r_{_\mathrm H}(1-\frac{\Lambda a^2}{3}-\Lambda{r_{_\mathrm H}^2}-\frac{a^2+q^2}{r_{_\mathrm H}^2})}{2(r_{_\mathrm H}^2+a^2)}\,,\qquad \Omega_{_\mathrm H}=\frac{a(1-\frac{\Lambda r_{_\mathrm H}^2}{3})}{r_{_\mathrm H}^2+a^2} \,, \qquad \Phi_{_\mathrm{H}}=\frac{qr_{_\mathrm H}}{r_{_\mathrm H}^2+a^2}.  
\end{equation}
By the choice of $\eta_{_\mathrm{H}}=\frac{2\pi}{\kappa_{_\mathrm{H}}}\{\zeta_{_\mathrm H},-\Phi_{_\mathrm{H}}\}$ in which $\zeta_{_\mathrm H}=\partial_t+\Omega_{_\mathrm H}\partial_\varphi$, the entropy variation can be found to be  
\begin{equation}
\hat \delta S_{_\mathrm{H}}\equiv \hat \delta H_{\eta_{_\mathrm{H}}}=\frac{\partial \left(\frac{\pi(r_{_\mathrm{H}}^2+a^2)}{G\,\Xi}\right)}{\partial m}\delta m + \frac{\partial \left(\frac{\pi(r_{_\mathrm{H}}^2+a^2)}{G\,\Xi}\right)}{\partial a}\delta a+ \frac{\partial \left(\frac{\pi(r_{_\mathrm{H}}^2+a^2)}{G\,\Xi}\right)}{\partial q}\delta q=\hat \delta \left(\frac{\pi(r_{_\mathrm{H}}^2+a^2)}{G\,\Xi}\right)\,,
\end{equation}
in which $\partial \Sigma$ can be chosen any $2$-dimensional spacelike smooth surface surrounding the singularity. Notice that $\eta_{_\mathrm{H}}$ contains a nonzero gauge transformation, which is almost fixed by the integrability condition. This gauge transformation provides a democratic picture in contribution of axial and gauge $U(1)$ symmetries to the entropy. Finally, by integration over solution phase space and choosing appropriate reference points, 
\begin{equation}
S_{_\mathrm{H}}=\frac{\pi(r_{_\mathrm{H}}^2+a^2)}{G\,\Xi}\,.
\end{equation}
\paragraph{First law(s):} Using the decomposition 
\begin{equation}
\eta_{_\mathrm{H}}=\frac{2\pi}{\kappa_{_\mathrm H}}\{\partial_t,0\}+\frac{2\pi\Omega_{_\mathrm H}}{\kappa_{_\mathrm H}}\{\partial_\varphi,0\}-\frac{2\pi\Phi_{_\mathrm{H}}}{\kappa_{_\mathrm H}}\{0,1\}=\frac{2\pi}{\kappa_{_\mathrm H}}\eta_{_M}+\frac{2\pi\Omega_{_\mathrm H}}{\kappa_{_\mathrm H}}\eta_{_J}-\frac{2\pi\Phi_{_\mathrm{H}}}{\kappa_{_\mathrm H}}\eta_{_Q}\,,
\end{equation}
by the linearity of $\hat \delta H_\eta$ in $\eta$, and $T_{_\mathrm{H}}=\frac{\kappa_{_\mathrm H}}{2\pi}$, the first laws would simply follow as
\begin{equation}
 \delta M=T_{_\mathrm{H}}\delta S_{_\mathrm{H}}+\Omega_{_\mathrm H}\delta J+\Phi_{_\mathrm{H}}\delta Q\,.
 \end{equation} 
At the end, we emphasize that in the analysis above, one can replace $r_{_\mathrm{H}}$ by the radius of each one of the horizons (including cosmological horizon), and the analysis would remain valid. But, one might use the standard choice in the BH context, \emph{i.e.} the BH event horizon.

\section{Conclusion}\label{sec-conclude}
Equipped with the SPSM, we analysed the thermodynamics of the Kerr-(Newman)-dS BHs, resulting in a coherent thermodynamic description which was reported in the Introduction,  and with the main features summarized below.
\begin{itemize}
\item \textbf{Democracy for the surfaces of integration:} Mass, angular momentum, and electric charge are charges attributed to the geometry, not any specific horizon or asymptotics. They can be calculated on any closed and smooth spacelike codimension-2 surface surrounding the singularity of the BH. Other conserved charges (\emph{e.g.} the entropies) respect this democracy too. 
\item \textbf{Unambiguous and regular conserved charges:} The mass, angular momentum, electric charge, and entropies are conserved charges associated with exact symmetries. So, they are calculated unambiguously, irrespective to the ambiguities in the symplectic structure. Moreover, the results are regular and finite automatically.
\item \textbf{Importance of the asymptotically non-rotating frame:} Mass is the conserved charge associated with the Killing vector $\partial_t$ in asymptotically non-rotating frame. Besides, the angular velocities which are manifest in the first law(s) are angular velocities in this frame.
\item \textbf{Democracy for the horizons:} For each one of the horizons, either cosmological or BH event horizons, one can associate temperature, angular velocity, electric potential, and entropy. The entropies as  conserved charges can be calculated on any spacelike, smooth, and closed codimension-2 surface surrounding the singularity of the BH. Moreover, for each one of the horizons there is an identity relating variations of mass, angular momentum, and electric charge to the variation of associated entropy, similar to the first law of thermodynamics.  
\item \textbf{Democracy for the signs of the cosmological constant:} Keeping the cosmological constant $\Lambda$ as an abstract parameter of the theory in the solution, the thermodynamics of Kerr, Kerr-AdS and Kerr-dS can be cast in a unified presentation.  
\end{itemize}  
Among the features above, the unambiguity and independence from the surfaces of integration have rigorous derivations. But, other features can be considered as some observations, which studying their physical and mathematical origins can be some interesting lines of research.  
\paragraph{Acknowledgement:} I would like to thank Shahin Sheikh-Jabbari for his crucial contributions to this paper, in addition to helpful discussions on the subject. I would also like to thank Erfan Esmaeili, who motivated me to study Kerr-dS BHs using SPSM. Besides, I thank Ali Seraj for all of the things he has taught me about covariant phase space formulation. This work has been supported by the \emph{Allameh Tabatabaii} Prize Grant of \emph{National Elites Foundation} of Iran and the \emph{Saramadan grant} of the Iranian vice presidency in science and technology.

\appendix
\section{A deeper review on solution phase space method}\label{app-deep}
The goal of this appendix is to provide a conceptual review on SPSM, although reference to the original paper \cite{HS:2015xlp} is recommended. Before reviewing SPSM, we need to recap a standard phase space construction, dubbed as \emph{covariant phase space formulation} \cite{Crnkovic1987,Ashtekar:1987hia,Ashtekar:1990gc,Lee:1990gr,Wald:1993nt,Iyer:1994ys}.  
\paragraph{Covariant phase space formulation:} Phase space $\mathcal{F}(\mathcal{M},\Omega)$ is a manifold $\mathcal{M}$ equipped with a closed nondegenerate  symplectic form $\Omega$. In classical field mechanics, it is usual to build the phase space canonically, \emph{i.e.}    building the $\mathcal{M}$ from a subset of field configurations $\Phi(\vec{x})$ and their momentum conjugates defined on some privileged time foliation of spacetime. In this construction, solutions to the equation of motion are some \emph{curves} on $\mathcal{M}$ parametrized by the time. Interestingly, in the context of generally covariant gravitational theories, there is a more suitable construction which does not break general covariance by specifying a time foliation. In this construction, $\mathcal{M}$ is composed of dynamical field configurations all over the spacetime $\Phi(x^\mu)$. On the other hand, there would not be any field conjugate present. As a result, any solution to the equation of motion in the phase space would be a \emph{point} on $\mathcal{M}$, instead of a curve. The tangent space of the manifold is also constituted from  a subset of perturbations $\delta \Phi(x^\mu)$. The symplectic $2$-form which makes $\mathcal{M}$ to be a phase space is constructed from the Lagrangian $d$-form $\mathbf{L}$. To this end, picking up the Lee-Wald $(d\!-\!1)$-form $\mathbf{\Theta}$ from the variation of Lagrangian
\begin{equation}\label{find Theta}
\delta \mathbf{L}=\mathbf{E}_{{\Phi}}\delta \Phi+\mrd \mathbf{\Theta}_{_\text{LW}}(\delta \Phi,\Phi)\,,
\end{equation}
the symplectic form would be \cite{Lee:1990gr,Wald:1993nt,Iyer:1994ys}
\begin{equation}\label{Omega LW}
\Omega_{_\text{LW}}(\delta_1\Phi,\delta_2\Phi,\Phi)\equiv \int_\Sigma \boldsymbol{\omega}_{_\text{LW}}(\delta_1\Phi,\delta_2\Phi,\Phi)\, 
\end{equation}
where
\begin{equation}\label{omega LW}
\boldsymbol{\omega}_{_\text{LW}}(\delta_1\Phi,\delta_2\Phi,\Phi)=\delta_1\mathbf{\Theta}_{_\text{LW}}(\delta_2\Phi,\Phi)-\delta_2\mathbf{\Theta}_{_\text{LW}}(\delta_1\Phi,\Phi)\,.
\end{equation}      
The $\mathbf{E}_{{\Phi}}$ denotes equation of motion for the field $\Phi$, the $\Sigma$ is some codimension-1 (Cauchy) surface and $\delta_{1,2}\Phi$ are some members of the tangent space. The $\boldsymbol{\omega}_{_\text{LW}}$ is called (pre)symplectic current. Closed-ness of $\Omega$ is guaranteed by the definition \eqref{omega LW}. In order to make $\Omega_{_\text{LW}}$ independent of the choice of $\Sigma$, one needs $\mrd \boldsymbol{\omega}_{_\text{LW}}=0$ and flow of $\boldsymbol{\omega}_{_\text{LW}}$ out of the boundaries $\partial \Sigma$ vanish. The former is achieved if $\Phi$ and $\delta\Phi$ satisfy e.o.m and linearized e.o.m respectively. So, it is standard to request them from the beginning.  But achievement of the latter needs extra conditions, usually some boundary conditions on perturbations. An important thing to be mentioned in covariant phase space formulation is the ambiguity of addition an exact $(d-1)$-form $\mrd\mathbf{Y}(\delta \Phi,\Phi)$ to the $\mathbf{\Theta}_{_\text{LW}}(\delta \Phi,\Phi)$, \emph{i.e.}
\begin{equation}
\mathbf{\Theta}_{_\text{LW}}(\delta \Phi,\Phi)\to \mathbf{\Theta}(\delta \Phi,\Phi)=\mathbf{\Theta}_{_\text{LW}}(\delta \Phi,\Phi)+\mrd\mathbf{Y}(\delta \Phi,\Phi)
\end{equation}
This ambiguity entails corresponding ambiguities in the $\Omega$ defined above, through 
\begin{equation}\label{omega ambiguity}
\boldsymbol{\omega}(\delta_1\Phi,\delta_2\Phi,\Phi)\to \boldsymbol{\omega}(\delta_1\Phi,\delta_2\Phi,\Phi)+ \mrd \big(\delta_2 \mathbf{Y}(\delta_1 \Phi,\Phi)-\delta_1 \mathbf{Y}(\delta_2 \Phi,\Phi)\big)\,.
\end{equation}
Using the symplectic form, one can associate a Hamiltonian generator (interchangeably called conserved charge) to a diffeomorphism+gauge transformation $\epsilon=\{\xi,\lambda^a\}$ as
\begin{align}\label{delta H xi}
\delta H_{\epsilon}(\Phi)&\equiv  \int_\Sigma \big(\delta^{[\Phi]}\mathbf{\Theta}(\delta_\epsilon\Phi,\Phi)-\delta_\epsilon\mathbf{\Theta}(\delta\Phi,\Phi)\big)=\int_{\Sigma}\mrd\boldsymbol{k}_{\epsilon}(\delta\Phi,\Phi)=\oint_{\partial\Sigma}\boldsymbol{k}_{\epsilon}(\delta\Phi,\Phi) \,. 
\end{align} 
The $\delta^{[\Phi]}$ emphasizes that $\delta$ acts on dynamical fields, not the $\epsilon$. Moreover, $\delta_\epsilon\Phi\equiv \mathscr{L}_\xi\Phi+\delta_{\lambda^a}A^a$ where $A^a$ are some probable Abelian gauge fields. In the equation above, the integrand in the first integration has been replaced by an exact $(d\!-\!1)$-form $\mrd \boldsymbol{k}_\epsilon$. So, the last equation follows from the Stokes theorem. The $(d\!-\!2)$-form $\boldsymbol{k}_\epsilon$ is explicitly as (see Appendix A in Ref. \cite{HS:2015xlp} for detailed derivation)
\begin{equation}\label{k_xi general}
\boldsymbol{k}_\epsilon(\delta\Phi,\Phi)=\delta \mathbf{Q}_\epsilon-\xi \cdot \mathbf{\Theta}(\delta \Phi,\Phi)\,,
\end{equation} 
in which $\mathbf{Q}_\epsilon$ is the \emph{Noether-Wald charge density}, defined by the relation
\begin{equation}\label{Noether Wald}
\mrd \mathbf{Q}_\epsilon\equiv \mathbf{\Theta}(\delta_\epsilon\Phi,\Phi)-\xi \! \cdot \! \mathbf{L}\,.
\end{equation}
Hence, by the Eq. \eqref{k_xi general},  $\boldsymbol{k}_\epsilon$ can be found for different theories straightforwardly. Putting it into Eq. \eqref{delta H xi}, if the last integral would be finite and nonvanishing,  $\delta H_{\epsilon}(\Phi)$ then corresponds to a conserved charge variation. In order to find the finite conserved charge $H_{\epsilon}$, integrability over the phase space is needed. This condition is basically $(\delta_1\delta_2-\delta_2\delta_1)H_\epsilon (\Phi)=0$, in which $\Phi$s are any field configuration in the presumed phase space $\mathcal{F}$, and $\delta_{1,2}\Phi$ are any arbitrary chosen member of its tangent space. Then, it follows that the integrability condition can be explained as \cite{Lee:1990gr,Wald:1999wa,Compere:2015knw}
\begin{equation}\label{integrability cond modified}
\oint_{\partial\Sigma} \Big(\xi\cdot \boldsymbol{\omega}(\delta_1\Phi,\delta_2\Phi,\Phi)+\boldsymbol{k}_{\delta_1\epsilon}(\delta_2\Phi,\Phi) -\boldsymbol{k}_{\delta_2\epsilon}(\delta_1\Phi,\Phi)\Big)=0.
\end{equation}

As far as calculation of conserved charges are concerned, conservation of $\delta H_\epsilon$ can be guaranteed if $\epsilon$ is chosen such that $\boldsymbol{\omega}(\delta\Phi,\delta_\epsilon\Phi,\Phi)=0$ on-shell. It is because there would not be any flow out of the boundaries locally, and hence globally. The family of $\epsilon$'s with this property, which has been dubbed ``symplectic symmetry generators" \cite{CHSS:2015mza}, can be divided to two sets: 1) the ones for which $\delta_\epsilon \Phi\neq0$ at least on one of the points of the phase space, 2) the ones for which $\delta_\epsilon \Phi=0$ all over the phase space. The former set, dubbed as ``nonexact symmetry generators", constitute a closed algebraic structure, and are considered to be responsible for generating the phase space of a solution at given constant thermodynamical variables. We can dub the generated phase space as ``statistical phase space". Hence, they open a road towards understanding microstates of the system (see \cite{CHSS:2015mza,CHSS:2015bca,Compere:2015knw} for works in this direction). The latter set are dubbed ``exact symmetry generators" and are considered as generators of the set of solutions in different thermodynamical variables \cite{HS:2015xlp}. The generated phase space has been called ``solution phase space" which we describe below. It has been conjectured that the phase space associated with the geometries without propagating degrees of freedom are composed of the combination of statistical and solution phase spaces \cite{HS:2015xlp}.
\paragraph{Solution phase space method:} This method is specification of the covariant phase space formulation to some specific manifolds and their tangent spaces which endows that method the power of calculability. Consider a family of (black hole) solutions to a generally covariant gravitational theory. Usually, such a family is identified by some isometries and some parameters $p_j$. The parameters are some arbitrary (but with constrained domain) real numbers appearing in the field configuration of the mentioned solutions. The parameters can be reparametrized, but can not be removed by coordinate transformations. The manifold $\hat{\mathcal{M}}$ can be chosen to be composed of the members of the family, up to unphysical coordinate/gauge transformations. The symplectic $2$-form $\hat \Omega$ would be simply the Lee-Wald symplectic form confined to $\hat{\mathcal{M}}$. Then, the $\mathcal{F}_p=(\hat{\mathcal{M}},\hat \Omega)$ would be a phase space, the ``solution phase space". Hence, any point of the manifold can be identified by $\hat \Phi(x^\mu,p_j)$. Tangent space of the $\hat{\mathcal{M}}$ is spanned (up to infinitesimal pure gauge transformations) by ``parametric variations" which are found simply by \cite{HSS:2014twa}
\begin{equation}
\hat{\delta}\Phi\equiv \frac{\partial \hat{\Phi}}{\partial p_j}\delta p_j.
\end{equation} 
These variations, which are infinitesimal difference of two solutions, satisfy linearized equation of motion. Hence, they respect $\mrd\boldsymbol{\omega}_{_\text{LW}}(\hat\delta_1\Phi,\hat\delta_2\Phi,\hat\Phi)=0$. 

As it was advertised above, conservation of $\hat\delta H_\epsilon$ is guaranteed if $\epsilon$ is chosen to be an exact symmetry generators $\eta$ defined in Eq. \eqref{exact symmetry}. This results is because of   $\boldsymbol{\omega}_{_\text{LW}}(\hat\delta\Phi,\delta_\eta\hat\Phi,\hat\Phi)=0$, (which itself is a result of linearity  of $\boldsymbol{\omega}_{_\text{LW}}$ in $\delta_\eta\hat\Phi=0$), preventing flow of $\boldsymbol{\omega}_{_\text{LW}}$ out of the boundaries $\partial \Sigma$. Along with guaranteeing the conservation, the relation $\boldsymbol{\omega}_{_\text{LW}}(\hat\delta\Phi,\delta_\eta\hat\Phi,\hat\Phi)=0$ yields an additional interesting and unexpected result:   $\hat\delta H_\eta$ would also be independent of the chosen $\partial\Sigma$. It is because of vanishing of $\boldsymbol{\omega}_{_\text{LW}}$  all over the $\Sigma$, and hence, vanishing of $\boldsymbol{\omega}_{_\text{LW}}$ in the region enclosed between two different integrating surfaces $\partial\Sigma_1$ and $\partial\Sigma_2$. Then, by the Stokes theorem, and noticing the Eq. \eqref{delta H xi}, the claim is proved.  Explaining  this result in another way, although the integration in calculating $\hat\delta H_\eta$ is over codimension-2 surface $\partial \Sigma$, but the result would be independent of all coordinates, including the two coordinates which are not integrated on.

Focusing on exact symmetries results in another nice feature for calculation of their conserved charges; discarding the ambiguity $\mathbf{Y}$. This is because of $\delta\mathbf{Y}(\delta_\eta\Phi,\Phi)-\delta_\eta\mathbf{Y}(\delta\Phi,\Phi)=0$, which is a result of the linearity  of the left hand side in $\delta_\eta\Phi=0$. Using this identity together with  Eq. \eqref{omega ambiguity} in the \eqref{delta H xi}, then there would not be any ambiguity in the definition of conserved charges as far as exact symmetries are concerned. Summarizing the last two paragraphs, the charges associated with exact symmetries are {conserved}, {unambiguous}, and independent of the chosen described surfaces of integration $\partial \Sigma$. 

So far, the SPSM has provided all materials needed to {calculate} $\hat \delta H_\eta (p_j)$. The final tasks are checking integrability over $\hat{\mathcal{M}}$, and (if integrable) performing the integration. The former is feasible simply by replacing $\delta\Phi$ and $\epsilon$ in Eq. \eqref{integrability cond modified} by $\hat \delta \Phi$ and $\eta$. The latter is  abstractly the integration in Eq. \eqref{finite H xi}, and pragmatically integrating $\hat \delta H_\eta (p_j)$ over the parameters $p_j$.

\paragraph{$\boldsymbol{k}_{\boldsymbol{\xi}}$ for EH-$\mathbf{\Lambda}$ theory:} To make the paper self-contained, here we provide the derivation of $\boldsymbol{k}_\xi$ for the EH-$\Lambda$ theory, which is described by the Lagrangian density $\mathcal{L}=\frac{1}{16\pi G} (R-2\Lambda)$. Beginning from the Eq. \eqref{find Theta}, one finds
\begin{align}\label{k EH proof 4}
\mathbf{\Theta}(\delta \Phi,\Phi)=\star\Big(\frac{1}{16\pi G}(\nabla_\alpha \delta g_{\,\,\mu}^{\alpha}-\nabla_\mu\delta g^\alpha_{\,\,\alpha})\,\mrd x^\mu\Big)\,.
\end{align}
In order to find the explicit form of the $\boldsymbol{k}_\xi$ through Eq.\eqref{k_xi general}, in addition to the equation above, the calculation of $\delta \mathbf{Q}_\xi$ is also needed. To this end, by the definition \eqref{Noether Wald} and using the equations of motion, 
\begin{align}
\mathbf{Q}_\xi&=\star \Big( \frac{-1}{16\pi G }\frac{1}{2!} (\nabla_\mu\xi_\nu-\nabla_\nu\xi_\mu)\,\mrd x^\mu\wedge\mrd x^\nu\Big)\\
&=\frac{-1}{16\pi G}\frac{\sqrt{-g}}{(2!(d-2)!)}\epsilon_{\mu\nu\alpha_1\dots\alpha_{d-2}}(\nabla^\mu\xi^\nu-\nabla^\nu\xi^\mu)\,\mrd x^{\alpha_1}\wedge\dots\wedge\mrd x^{\alpha_{d-2}}\,.\label{k EH proof 2}
\end{align}
 Now by the relations 
\begin{align}
\delta \sqrt{-g}=\frac{\sqrt{-g}}{2}\delta g^\alpha_{\,\,\alpha}\,, \qquad \delta \Gamma ^\lambda_{\mu\nu}&= \frac{1}{2}[g^{\lambda \sigma}\big(\nabla _\mu\delta  g_{\sigma \nu}+\nabla_\nu\delta g_{\sigma \mu}-\nabla _\sigma \delta g_{\mu\nu}\big)]\,,
\end{align}
one finds
\begin{align}
\delta \mathbf{Q}_\xi&=\frac{-1}{16\pi G}\frac{\sqrt{-g}}{(2!(d-2)!)}\epsilon_{\mu\nu\alpha_1\dots\alpha_{d-2}}\Big(\frac{1}{2}\delta g^\alpha_{\,\,\alpha}(\nabla^\mu\xi^\nu)-\delta g^{\mu\beta}(\nabla_\beta\xi^\nu)\nonumber\\
&\hspace*{4cm}+\xi^\alpha\nabla^\mu \delta g^\nu_{\,\,\alpha}\Big)\,\mrd x^{\alpha_1}\wedge\dots\wedge\mrd x^{\alpha_{d-2}}-[\mu\leftrightarrow\nu]\,.\label{k EH proof 5}
\end{align}
in which the notation $\delta g^{\mu\nu}\equiv g^{\mu\alpha}g^{\nu\beta}\delta g_{\alpha\beta}=-\delta (g^{\mu\nu})$ has been used. Notice that by $\delta (g^{\mu\nu})$ we meant the direct action of $\delta$ on $g^{\mu\nu}$. The next step in calculating the $\boldsymbol{k}_\xi$ would be finding the second term in \eqref{k_xi general}, which is

\vspace*{-0.3cm}
\small{
\begin{align}
-\xi \cdot \mathbf{\Theta}(\delta \Phi,\Phi)&=-\xi \cdot\Big(\frac{1}{16\pi G}\frac{\sqrt{-g}}{(d-1)!}\epsilon_{\mu\alpha_{1}\dots\alpha_{d-1}}(\nabla_\alpha \delta g^{\alpha\mu}-\nabla^\mu\delta g^\alpha_{\,\,\alpha})\mrd x^{\alpha_1}\wedge\dots \wedge \mrd x^{\alpha_{d-1}}\Big)\nonumber\\
&=\frac{-1}{16\pi G}\frac{\sqrt{-g}}{(d-2)!}\epsilon_{\mu\nu\alpha_{1}\dots\alpha_{d-2}}(\nabla_\alpha \delta g^{\alpha\mu}-\nabla^\mu\delta g^\alpha_{\,\,\alpha})\xi^\nu \,\mrd x^{\alpha_1}\wedge\dots \wedge \mrd x^{\alpha_{d-2}}\\
&=\frac{-1}{16\pi G}\frac{\sqrt{-g}}{2(d-2)!}\epsilon_{\mu\nu\alpha_{1}\dots\alpha_{d-2}}(\nabla_\alpha \delta g^{\alpha\mu}-\nabla^\mu\delta g^\alpha_{\,\,\alpha})\xi^\nu \,\mrd x^{\alpha_1}\wedge\dots \wedge \mrd x^{\alpha_{d-2}}-[\mu\leftrightarrow\nu]\,.\label{k EH proof 6}
\end{align}}\normalsize

\vspace*{-0.5cm}
\noindent Finally, having found the \eqref{k EH proof 5} and \eqref{k EH proof 6}, the $\boldsymbol{k}_\xi^{\text{EH}}$ can be read as
\begin{align}
\boldsymbol{k}_\xi^{\text{EH}}&=\frac{-1}{16\pi G}\frac{\sqrt{-g}}{(2!(d-2)!)}\epsilon_{\mu\nu\alpha_1\dots\alpha_{d-2}}\Big(\frac{1}{2}\delta g^\alpha_{\,\,\alpha}(\nabla^\mu\xi^\nu)-\delta g^{\mu\beta}(\nabla_\beta\xi^\nu)+\xi^\alpha\nabla^\mu \delta g^\nu_{\,\,\alpha}\nonumber\\
&+(\nabla_\alpha \delta g^{\alpha\mu}-\nabla^\mu\delta g^\alpha_{\,\,\alpha})\xi^\nu\Big)\,\mrd x^{\alpha_1}\wedge\dots\wedge\mrd x^{\alpha_{d-2}}-[\mu\leftrightarrow\nu]\,.
\end{align}
By the Hodge duality, we would have $\boldsymbol{k}_\xi^{\text{EH}}=\star k_\xi^{\text{EH}}$, where
\begin{equation}
k_\xi^{\text{EH}\mu\nu}=\frac{-1}{16\pi G}\Big(\frac{1}{2}\delta g^\alpha_{\,\,\alpha}(\nabla^\mu\xi^\nu)-\delta g^{\mu\beta}(\nabla_\beta\xi^\nu)+\xi^\alpha\nabla^\mu \delta g^\nu_{\,\,\alpha}+(\nabla_\alpha \delta g^{\alpha\mu}-\nabla^\mu\delta g^\alpha_{\,\,\alpha})\xi^\nu\Big)-[\mu\leftrightarrow\nu]\,.
\end{equation}
Notice that this result is independent of the cosmological constant $\Lambda$.


\end{document}